\documentclass[aps,prb,twocolumn,showpacs,amsmath,amssymb]{revtex4}

\usepackage{graphicx}
\usepackage{amssymb}
\usepackage{dcolumn}
\usepackage{bm}
\usepackage{epstopdf}
\usepackage{float,epsfig}
\newcommand{\dpart}[2]{\dfrac{\partial#1}{\partial#2}}
\newcommand{\ddiff}[2]{\dfrac{\mathrm{d}#1}{\mathrm{d}#2}}

\begin{document}

\title{Theory of I-V Characteristics of Magnetic Josephson Junctions}
\author{Shengyuan A. Yang}
\affiliation{Department of Physics, The University of Texas, Austin,
Texas, 78712-0264, USA}

\author{Qian Niu}
\affiliation{Department of Physics, The University of Texas, Austin,
Texas, 78712-0264, USA}

\author{D. A. Pesin}
\affiliation{Department of Physics, The University of Texas, Austin,
Texas, 78712-0264, USA}

\author{A. H. MacDonald}
\affiliation{Department of Physics, The University of Texas, Austin,
Texas, 78712-0264, USA}

\begin{abstract}
We analyze the electrical characteristics of a circuit consisting of
a free thin-film magnetic layer and source and drain electrodes that
have opposite magnetization orientations along the free magnet's two
hard directions.  We find that when the circuit's current exceeds a
critical value there is a sudden resistance increase which can be
large in relative terms if the currents to source or drain are
strongly spin polarized and the free magnet is thin.  This behavior
can be partly understood in terms of a close analogy between the
magnetic circuit and a Josephson junction.
\end{abstract}

\pacs{85.75.-d 75.47.-m 75.75.-c 72.25.-b}

\maketitle

\section{Introduction}
Electronic transport can usually be described in terms of
effectively independent electrons. Recently, with the discovery and
exploitation of spin-transfer
torque\cite{slonczewski_stt,berger_stt,tsoi_stt,ralph_stt} (STT)
effects, magnetism has joined superconductivity as an instance in
which collective and quasiparticle contributions to transport are
entwined in an essential way.  The similarity between the
non-equilibrium properties of magnetic and
superconducting\cite{nonequilsuper} systems is especially close when
comparing the properties of a superconducting circuit containing a
Josephson junction to a magnetic circuit containing a free
ferromagnetic layer with strong easy-plane anisotropy.  As we
explain in more detail below, the role of the Josephson junction
bias current in the superconducting circuit is played in the
magnetic case by the component of the spin-current injected into the
nanoparticle that is perpendicular to the easy plane.

\begin{figure}
\centering \epsfig{file=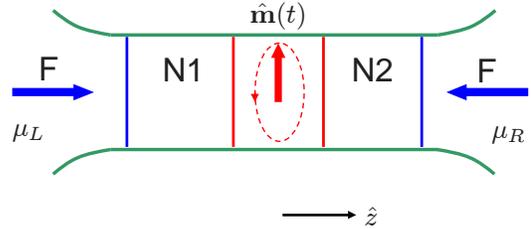,width=0.8\linewidth}
\caption{\label{fig:one} (color online). A magnetic Josephson
junction consists of a free magnetic layer with a well-defined
easy-plane and weak in-plane anisotropy separated by non-magnetic
spacers from magnetic source and drain electrodes whose
magnetizations are opposite and perpendicular to the easy plane. In
this article we choose the spin $\hat{z}$-axis along a free magnet
hard direction.}
\end{figure}

The electrical properties of a circuit containing a Josephson
junction typically change drastically when the junction's critical
current is exceeded.  In this paper we propose that the magnetic
circuit illustrated in Fig.~\ref{fig:one}, which we refer to as a
magnetic Josephson junction (MJJ), can exhibit similar drastic
effects when a critical current related to the free magnet's
in-plane anisotropy is exceeded. We show that the resistance of an
MJJ can increase substantially when its critical current is exceeded
provided that either source or drain currents are strongly spin
polarized and magnetization relaxation in the free magnet layer is
not too rapid.

The analogy we explore is closely related to the early suggestion by
Berger\cite{berger_JJ} that 180$^{\circ}$ domain walls in
ferromagnets should exhibit behavior analogous to the behavior of
Josephson junctions.  Indeed the MJJ geometry we propose may be
viewed as a discrete version of a pinned 180$^{\circ}$ domain wall.
Although the magnetization dynamics induced {\em emf} that Berger
predicted based on an analogy to the {\em ac} Josephson relation has
now\cite{beach_1,beach_2} been confirmed experimentally, electrical
signals of magnetization texture dynamics in uniform bulk materials
tend to be weak. The MJJ geometry we study is also closely related
to that employed in spin-transfer torque
oscillators.~\cite{silva_sto} It is well known that the dc
resistance of an STT-oscillator tends to increase once the
magnetization becomes dynamic.~\cite{tsoi_stt}  The increase in bias
voltage at fixed current is closely related to Berger's Josephson
voltage. From this point of view, the issue we explore theoretically
in this paper is the possibility of achieving large relative changes
of resistance in an STT-oscillator when the magnetization becomes
dynamic.  We believe that such an examination is motivated by recent
advances in ferromagnetic metal
spintronics~\cite{perpendicular,strongly} which have enabled the
injection of more highly spin polarized currents and decreased the
thickness of free magnetic layers, making them easier to manipulate
electrically.   MJJ devices not only provide a test bed for our
understanding of the fundamental physics of spin gauge fields, but
could also be useful because of their unusual transport properties.

Our paper is organized as following. In Sec. II, we comment at
greater length on the relationship between Josephson junctions and
easy-plane free magnets. In Sec. III, we discuss the theoretical
framework we used for analyzing the transport properties of an MJJ.
In Sec. IV and Sec. V, we identify two regimes in which the MJJ
could have quite different resistances. Finally, in Sec. VI, we
summarize our results and discuss our conclusions.

\section{Nanomagnet Equations of Motion}
As shown in Fig.~\ref{fig:one}, the MJJ is a multilayer
heterostructure consisting of ferromagnetic and non-magnetic normal
metal layers. The two ferromagnetic electrodes have opposite
magnetization directions, serving as spin polarized current source
and drain. The free magnetic layer in the middle is sandwiched
between two non-magnetic normal metal spacer layers to avoid direct
magnetic coupling with the electrodes. We assume that the free
magnetic layer is small enough that its magnetization is spatially
constant. Its magnetization direction dynamics is then described by
a Landau-Liftshitz-Gilbert (LLG) equation,
\begin{equation}
\ddiff{\hat{\textbf{m}}}{t}=-\gamma \, \hat{\textbf{m}}\times\textbf{H}_\text{eff}+\alpha \,
\hat{\textbf{m}}\times\ddiff{\hat{\textbf{m}}}{t}+\left.\ddiff{\hat{\textbf{m}}}{t}\right|_\text{torque},
\end{equation}
where $\hat{\textbf{m}}$ is a unit vector along the magnetization
direction of the free layer, $\gamma=|g|\mu_B/\hbar>0$ is (minus)
the gyromagnetic ratio, $\textbf{H}_\text{eff}$ is an effective
magnetic field, $\alpha$ is the dimensionless Gilbert damping
constant and the right hand size of the equation above denotes a
Slonczewski\cite{slonczewski_stt,berger_stt,tsoi_stt,ralph_stt}
spin-transfer torque term\cite{spintransfercaveat} that captures the
coupling between transport and collective dynamics:
\begin{equation}
\left.\ddiff{\hat{\textbf{m}}}{t}\right|_\text{torque}=\frac{1}{S}
\; (\hat{\textbf{m}}\times \textbf{I}_s)\times\hat{\textbf{m}}.
\end{equation}
Here $S=M_sV/\gamma$ is the magnitude of the total spin of the free
magnetic layer, $V$ is the free layer volume, $M_s$ is its
magnetization per unit volume, $\textbf{I}_s$ is the net spin
current flowing out of the free layer and $(\hat{\textbf{m}}\times
\textbf{I}_s)\times\hat{\textbf{m}}=\textbf{I}_s-(\textbf{I}_s\cdot\hat{\textbf{m}})\hat{\textbf{m}}$
selects the component of $\textbf{I}_s$ that is transverse to
$\hat{\textbf{m}}$. We assume that there is no applied magnetic
field. The effective magnetic field,
\begin{equation}
\textbf{H}_\text{eff}=-\frac{1}{\gamma S}\frac{\delta F}{\delta
\hat{\textbf{m}}},
\end{equation}
then arises mainly from the magnetostatic energy of a thin film with
an elliptical cross-section:
\begin{equation}
F=K_z \, m_z^2+K_p \, \sin^2\phi,
\end{equation}
where $\phi$ is the magnetization orientation azimuthal angle and
the anisotropy parameter $K_z \gg K_p$ so that the magnetization
direction is always close to its easy plane. When the magnetization
direction is expressed in terms of $m_z$ and $\phi$, and $m_z$ is
assumed to be small, the LLG equations take the form
\begin{equation}\label{mzdot}
\dot{m}_z=\frac{1}{S}\dpart{F}{\phi}+\alpha \dot{\phi}+\frac{1}{S}\,
I^z_s,
\end{equation}
\begin{equation}\label{phidot}
\dot{\phi}=-\frac{1}{S}\dpart{F}{m_z}-\alpha \dot{m}_z+ \frac{1}{S} \,
I_s^\phi.
\end{equation}
As discussed later we will be most interested in circumstances under
which $\dot{m}_z=0$ and $I_s^\phi$, the component of spin-current in
the azimuthal direction, vanishes. In that limit the right hand side
of the equation for $\dot{\phi}$ reduces to the difference between
the chemical potential of majority and minority spin electrons.
Eq.~(\ref{mzdot}) is then precisely equivalent to the resistively
and capacitively shunted Josephson junction model\cite{tinkham} with
easy-plane anisotropy playing the role of the capacitance,
\begin{equation}
\dot{m}_{z} = -\frac{S}{2 K_{z}} \; \frac{\mathrm{d}^2
\phi}{\mathrm{d} t^2},
\end{equation}
in-plane anisotropy playing the role of the superconducting coupling
which supports a supercurrent
\begin{equation}
I_\text{sup} =  K_{p} \sin(2\phi),
\end{equation}
$\alpha \dot{\phi}$ playing the role of a dissipative quasiparticle current, and
the spin-transfer torque term playing the role of junction bias current.

It is well known that the electrical properties of a Josephson
junction changes drastically at its critical current. Due to the
connections revealed above, we expect that similar behavior can
occur in an MJJ. Because the MJJ bias current is a spin-current not
a charge current, and its dissipationless supercurrent transports
spins from up to down states not charges from one side of the
junction to the other, the current-voltage relationship of a circuit
in which an MJJ is embedded is more subtle than in the
superconducting Josephson junction case, and collective effects are
generally weaker. In the following, we analyze MJJ I-V relationships
with the aim of identifying circumstances under which the resistance
change at the critical current can be large in circuits that are
fabricated with modern spintronic materials.

\section{Magnetoelectronic Circuit Theory}
For small bias voltages the nanoparticle magnetization can achieve a
static value by balancing the circuit bias and collective transport
terms: $I_s^z=I_\text{sup}$.  Once $I_\text{sup}$ reaches its
maximum value, $I_\text{crit}= K_p$, the magnetization's azimuthal
angle becomes time-dependent and the time average of $I_\text{sup}$
quickly becomes negligible.  In this dynamic limit $I_s^z$ must be
balanced against the dissipative term $-\alpha \dot{\phi}$. Our main
goal in this section is to shed light on the requirements for a
substantial relative change in circuit resistance when the MJJ's
critical current is exceeded. For this purpose we use
magnetoelectronic circuit
theory,\cite{Tserkovnyak_RMP,Brataas_PRL,Brataas_Review} which
simplifies the transport problem in ferromagnet/normal metal
heterostructures by dropping some non-local effects and separating
the circuit into normal metal and magnetic constituents within which
the exchange field is taken to be constant in direction. In the
following, we assume that the ferromagnetic layer is thicker than
the magnetic coherence length~\cite{Brataas_Review} which is on the
order of ${\AA}$ in transition metals.

Magnetoelectronic circuit theory's key equation relates the charge
and spin currents that flow from a ferromagnet into a normal metal
to the ferromagnet's chemical potential and to the chemical
potential of the normal metal, including its non-equilibrium
spin-accumulation contributions. The charge current
\begin{equation}
\label{Icharge}
I_c=\frac{e}{2h}\left[2(g^{\uparrow,\uparrow}
+g^{\downarrow,\downarrow})\; (\mu_{Nc}-\mu_F)+(g^{\uparrow,\uparrow}-g^{\downarrow,\downarrow}) \; \bm{\mu}_{Ns} \cdot
\hat{\textbf{m}}_{F} \right].
\end{equation}
Here $\uparrow$ and $\downarrow$ refer to majority and minority
spins, $\hat{\textbf{m}}_{F}$ is the ferromagnet's magnetization
direction, $g^{\sigma,\sigma'}$ is a dimensionless spin-dependent
conductance, $\mu_{Nc}$ is the normal metal chemical potential
averaged over spin states while $\bm{\mu}_{Ns}$ represents the
magnitude and direction in spin-space of the spin-dependent part of
the normal metal's chemical potential.  For static magnetization the
analogous spin-current expression\cite{gcaveat} is
\begin{widetext}
\begin{equation}
\label{Ispin} \textbf{I}_s =
-\frac{1}{8\pi}\left[2(g^{\uparrow,\uparrow}
-g^{\downarrow,\downarrow})(\mu_{Nc}-\mu_{F})\hat{\textbf{m}}_{F}+(g^{\uparrow,\uparrow}
+g^{\downarrow,\downarrow})(\bm{\mu}_{Ns} \cdot
\hat{\textbf{m}}_{F})\hat{\textbf{m}}_{F}+2g^{\uparrow,\downarrow}\hat{\textbf{m}}_{F}
\times\bm{\mu}_{Ns} \times\hat{\textbf{m}}_{F} \right].
\end{equation}
\end{widetext}
Eqs.~(\ref{Icharge},~\ref{Ispin}) apply to the four
ferromagnet/normal metal interfaces in a MJJ. We assume that the
normal metal nodes have sizes smaller than the spin flip diffusion
length, allowing its spin flip scattering to be neglected. Setting
the total charge and spin-currents into each of the two normal metal
nodes to zero yields eight equations for the eight unknown
parameters which specify the spin-dependent chemical potentials of
the two normal metal nodes. In the following two sections, we shall
calculate the resistance for both the static regime and the dynamic
regime from this set of equations.

\section{Static MJJ Resistance}
To apply magnetoelectric circuit theory to an MJJ with a
time-independent free magnet, we note that $\hat{\textbf{m}}_F = \pm
\hat{\bm{z}}$ for source and drain electrodes respectively, set the
source and drain chemical potentials to $\pm U/2$ and assume that
source and drain ferromagnetic electrodes are identical. It follows
from symmetry that spin-accumulation in the direction perpendicular
to $\hat{\textbf{m}}$ and $\hat{\bm{z}}$ vanishes in both normal
metal nodes, reducing the number of free variables to six. We
separate the normal metal chemical potentials into accumulation
($+$) and transport ($-$) contributions defined by
\begin{equation}
\mu_{\alpha,\pm} \equiv \mu_{\alpha,N1} \pm \mu_{\alpha,N2},
\end{equation}
where $\alpha=c,m,z$ for charge, free magnet, and source spin
directions, and $Ni$ $(i=1,2)$ denotes the $i$-th normal metal node.

Simple equations for $\mu_{\alpha,\pm}$ can be obtained by adding
and subtracting the charge and spin conservation conditions for the
two nodes. In this way it is easy to verify that for the static case
$\mu_{c,+}=\mu_{m,+}=\mu_{z,-}=0$, properties that also can be
established by symmetry considerations. The three remaining
equations determine the chemical potential differences that drive
charge and spin across the free magnetic layer, $\mu_{c,-}$ and
$\mu_{m,-}$, and the accumulation of spin injected by the source and
drain electrodes $\mu_{z,+}$:
\begin{eqnarray}\label{static}
2(g+2G) \mu_{c,-} + gp \mu_{z,+} + 2GP \mu_{m,-} &=& 2gU, \nonumber \\
2gp \mu_{c,-} + (g+G\eta) \mu_{z,+} &=& 2 gpU, \nonumber \\
4GP \mu_{c,-} + (2G+g\eta') \mu_{m,-} &=& 0,
\end{eqnarray}
where we have introduced the shorthand notations
$g=g^{\uparrow,\uparrow}+g^{\downarrow,\downarrow}$,
$p=(g^{\uparrow,\uparrow}-g^{\downarrow,\downarrow})/g$ and
$\eta'=2g^{\uparrow,\downarrow}/g$ for the interfaces between normal
metal and source and drain electrodes.  The corresponding upper case
$G$'s and $\eta$ refer to interfaces between the normal metals and
the free magnetic layer.

The charge current is most conveniently evaluated at either source
or drain electrode with the result that $I_c = (eg/2h)
(\mu_{c,-}+p\mu_{z,+}/2- U)$.  Although the static MJJ resistance
$R_S=U/(-eI_c)$ depends on the detailed values of the various
interface conductance parameters, it tends to be close to the sum of
the individual interface resistances and is not especially sensitive
to source or drain electrode spin-polarization; in the limit in
which all ferromagnets are fully polarized and identical, for
example, the static MJJ resistance is $5 (h/e^2)/g$. This result
applies until the perpendicular spin current $I^{z}_{s} =
(G\eta/8\pi) \; \mu_{z,+} $ reaches its critical value
$I_\text{crit}=K_p$. The corresponding critical charge current
flowing through the MJJ is proportional to $I_\text{crit}$. For
typical values $p\sim 1$, $\eta\sim 2$, $K_p\sim 10^5$J (per m$^3$)
and assuming that $G=g$, the citical charge current density is on
the order of $1\times 10^{-7}$A/nm$^2$ for a free layer with a
thickness around $1$nm.

\section{Dynamic MJJ Resistance}
When $I^{z}_s$ is substantially
larger than $I_\text{crit}=K_p$ a dynamic steady state is approached
in which
\begin{equation}\label{18}
\dot{\phi}=-\frac{1}{\alpha S} I^z_s.
\end{equation}
This precession of magnetization induced by spin current is the
working principle of STT-oscillators. Because of their long spin
relaxation times, the normal metal nodes do not follow this
precession so that the components of the spin-dependent part of the
chemical potentials vanish along directions other than
$\hat{\bm{z}}$. This leaves four unknown chemical potentials and
$\dot{\phi}$ to be determined by Eq.~(\ref{18}) and the four circuit
equations for conservation of $I_c$ and $I_z$ at each node. In the
dynamic case the total spin-current $I^z_s$ flowing out of the free
magnet includes a spin-pumping
contribution:~\cite{Tserkovnyak_RMP,Tserkovnyak_spinpumping}
\begin{equation}
I_s^z=-\frac{1}{8\pi}G \eta (\mu_{z,+}-2\hbar\dot{\phi}).
\end{equation}
The induced pumping current tends to counter the bias
current, effectively increasing the resistance.
Following the same
strategy as in the static case we find that $\mu_{c,+}=\mu_{z,-}=0$
while the remaining unknowns satisfy
\begin{eqnarray}
\label{dynamic}
2(g+2G) \mu_{c,-} + gp \mu_{z,+} &=& 2 g U,\nonumber \\
2 gp \mu_{c,-} +(g+G\eta) \mu_{z,+} - 2G\eta \hbar \dot{\phi} &=& 2 g p U, \nonumber \\
G \eta \mu_{z,+} -  2 G\eta \hbar \dot{\phi} &=& 8\pi \alpha S
\dot{\phi}.
\end{eqnarray}
The third of these equations relates the spin-current out of the
free nanoparticle to the spin decay rate.  The electrical properties
of a MJJ are most interesting when the dimensionless coupling
constant formed by comparing the right and left hand sides of
Eq.~(\ref{dynamic}),
\begin{equation}
\Lambda \equiv \frac{ 4\pi \alpha S}{\hbar G\eta},
\end{equation}
is small.  Taking typical values\cite{Tserkovnyak_RMP} of $G \sim
10$ (per nm$^{2}$), $\eta\sim 2$ and of $(4 \pi M_s)/\gamma \sim 100
\, {\rm nm}^{-3}$ yields $\Lambda \sim 10 \,\alpha\, d[\rm{nm}]$,
where $d$ is the film thickness, suggesting that small values are
possible in thin magnetic films made from materials like permalloy
that have small values of $\alpha$.  Solving Eqs.~(\ref{dynamic})
for $\Lambda=0$ we find that the dynamic MJJ resistance
\begin{equation}
R_D = \frac{h}{e^2} \;  \frac{4G + 2g(1-p^2)}{4Gg(1-p^2)},
\end{equation}
which diverges for $p \to 1$.  When source and drain electrodes are strongly spin-polarized and
$\Lambda$ is small, the MJJ circuit should suffer a large increase in resistance at its critical current.

This drastic resistance change can be qualitatively explained as
follows. In the static regime, the free layer magnetized in the
$xy$-plane breaks spin rotation symmetry along $z$-axis.
Exchange-field driven spin flips then provides a path for electrical
conduction between oppositely polarized electrodes.  In the dynamic
regime, precession of the free layer restores spin rotation symmetry
and the $z$-component of the conduction electron spin is
approximately conserved, increasing the electrical flow resistance.

\begin{figure}
\centering \epsfig{file=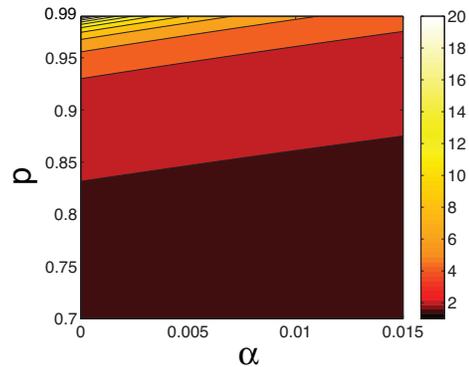,width=0.7\linewidth}
\caption{\label{fig:two} (color online). The ratio of dynamic and
static resistance $\zeta\equiv R_D/R_S$ plotted as a function of
damping parameter $\alpha$ and polarization $p$. The parameters used
here are $g=G=10$ (per nm$^2$), $P=0.5$, $\eta=\eta'=2$,
$M_s=10^5{\rm A/m}$, and $d=1{\rm nm}$. This ratio $\zeta$ diverges
as $\alpha \rightarrow 0$ and $p\rightarrow 1$. }
\end{figure}

In Fig.~\ref{fig:two}, we plot the ratio $\zeta\equiv R_D/R_S$,
which characterizes the performance of a MJJ, as a function of the
damping parameter $\alpha$ and the polarization $p$. $\zeta$
increases as $\alpha$ decreases or $p$ increases, and diverges in
the limit $\alpha \rightarrow 0$ and $p\rightarrow 1$.

\begin{figure}
\centering \epsfig{file=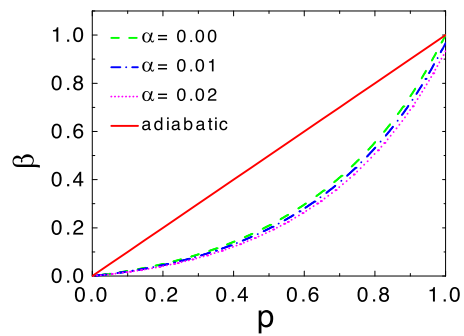,width=0.8\linewidth}
\caption{\label{fig:three} (color online). The ratio between $\hbar
\dot{\phi}/e$ and the magnetization dynamics induced \emph{emf} in a
MJJ plotted as a function of polarization $p$ for different values
of damping parameter $\alpha$.  These results were obtained for
$P=p$, $g=G=10$ (per nm$^2$), $\eta=\eta'=2$, $M_s=10^5{\rm A/m}$,
and $d=1{\rm nm}$. The straight line $\beta=p$ applies for smooth
magnetization textures.}
\end{figure}

\section{Discussion}
The influence of magnetization dynamics on electronic transport can
be described by accounting for spin pumping, as in the analysis
summarized above, or equivalently by accounting for a dynamics
induced \emph{emf}. For a slowly varying magnetic texture like a
domain wall in transition metal ferromagnets, a transport electron
follows the local magnetization direction adiabatically. The induced
\emph{emf} is\cite{beach_1,beach_2} then given by
$p\hbar\dot{\phi}/e$ with $\dot{\phi}$ being the domain wall
precession frequency and $p$ the bulk transport current
spin-polarization. For the MJJ considered here, spin transport is no
longer adiabatic. In order to relate the two perspectives we assume
that the influence of precession in the MJJ can also be captured by
imposing an extra \emph{emf} $\beta\hbar\dot{\phi}/e$ that accounts
for the difference between static and dynamic circuit resistances:
\begin{equation}\label{beta}
\frac{U}{R_D}=\frac{U-\beta\hbar\dot{\phi}}{R_S},
\end{equation}
The effective value of $\beta$ then depends on microscopic MJJ
parameters and can be evaluated by comparing Eq.~(\ref{beta}) with
the results of a magnetoelectronic circuit analysis. Some typical
values of $\beta$ are plotted in Fig.~\ref{fig:three} as a function
of $p$ ($=P$) for different values of $\alpha$. For the completely
adiabatic case, we should expect $\beta=p$ which is a straight line
in the figure.  Although the $\beta$ of a MJJ generally deviates
from this straight line due to nonadiabatic effects, it approaches
the straight line in the $p\rightarrow 0$ and $p\rightarrow 1$
limits.  For $p=1$, the same result can be obtained by making a
unitary transformation into the rotating frame of $\hat{\textbf{m}}$
and the result amounts to a relative shift $\hbar\dot{\phi}$ between
the chemical potentials of the two electrodes.  We find that the
magnetization dynamics damping, parameterized by $\alpha$, slightly
decreases the value of $\beta$ because the dissipation of energy
through damping can be regarded as an extra resistance in the
circuit.

From the analysis above, the current drops once bias voltage $U$
increases above a critical value determined by the in-plane
anisotropy, i.e. when the precession of magnetization begins. This
signals a negative resistance region of the I-V curve similar to
that for the Gunn diode.~\cite{hess} For MJJ with high performance
($\zeta\gg 1$), it could be used as a good fuse link, for example,
to control the maximum current that can flow through a circuit.

To summarize, we propose a ferromagnet/normal metal heterostructure
which has properties analogous to those of a Josephson junction. We
analyze its I-V characteristics and identify a static regime in
which the magnetization configuration is static with a low
resistance and a dynamic regime in which the magnetization precesses
and the resistance is high. Our analysis indicates that MJJ circuits
which exhibit large and sudden relative change in resistance can be
fabricated using modern spintronic materials with perpendicular
anisotropy\cite{perpendicular} and strongly polarized spin
currents\cite{strongly}. Besides having an academic interest,
because of the insight that it provides in comparing current-induced
emf and spin-pumping notions in metal spintronics, the MJJ may be
useful for applications.

\section{Acknowledgements}
The authors thank Tomas Jungwirth, Jairo Sinova and Maxim Tsoi for
helpful discussions. This work was supported by NSF (DMR0906025),
the Welch Foundation grant F1473, DOE (DE-FG02-02ER45958, Division
of Materials Science and Engineering) and Texas Advanced Research
Program.


\begin{references}
\bibitem{spintransfercaveat}  Besides Slonczewski spin-transfer term, which
captures the overall conservation of angular momentum, there can be other
contributions to the coupling between transport currents and magnetization dynamics.
See Ref.[\onlinecite{ralph_stt}].
\bibitem{gcaveat} We have in the interests of transparency dropped a small term
proportional to the imaginary part of the
spin-flip conductance that does not play an essential role in our discussion.
\end{references}
\end{document}